\documentclass[pra,aps,twocolumn,showpacs]{revtex4-1}
\usepackage{graphicx}
\usepackage{bm,color}
\usepackage{physics}
\usepackage{amsmath, amssymb}
\usepackage{bm,color}
\usepackage{multirow}
\usepackage{ulem}

\newcommand{\be}{\begin{eqnarray}}
\newcommand{\ee}{\end{eqnarray}}
\newcommand{\nn}{\nonumber\\}

\usepackage{ulem}

\newcommand{\cbin}[1]{{\color{blue}#1}}
\renewcommand{\cbin}[1]{}

\renewcommand{\theequation}{\arabic{equation}}

\begin{document}

\title{
Study of quantum non-locality by CHSH function and its extension \\in disordered fermions
}
\date{\today}
\author{Yoshihito Kuno$^{1}$}

\affiliation{$^1$Graduate School of Engineering Science, Akita University, Akita 010-8502, Japan}

\begin{abstract}
Quantum non-locality is an important concept in quantum physics. In this work, we study the quantum non-locality in a fermion many-body system under quasi-periodic disorders. The Clauser-Horne-Shimony-Holt (CHSH) inequality is systematically investigated, which quantifies quantum non-locality between two sites. 
We find that the quantum non-locality explicitly characterize the extended and critical phase transitions, and further that in the globally averaged picture of maximum value of the quantum non-locality the CHSH inequality is not broken, but for a local pair in the internal of the system the violation probability of the CHSH inequality becomes sufficiently finite. Further we investigate an extension of the CHSH inequality, Mermin-Klyshko-Svetlichny (MKS) polynomials, which can characterize multipartite quantum non-locality. We also find a similar behavior to the case of CHSH inequality. In particular, in the critical regime and on a transition point, the adjacent three qubit MKS polynomial in a portion of the system exhibits a quantum non-local violation regime with a finite probability. 
\end{abstract}


\maketitle
\section{Introduction}
Study of quantum correlation is one of the most attracted research field in physics \cite{Horodecki_family}. 
In particular, the characterization of quantum correlation for many body system is now on-going problem. 
Quantum entanglement is one of the most famous measure of quantum correlation in many-body system. The behaviors of quantum entanglement or entanglement entropy have been extensively studied \cite{Horodecki_family,Eisert2010}. 
They succeeded in characterizing non-trivial phases such as symmetry protected topological phase, topological ordered phase \cite{Wen_text}. 

There is another concept elucidating quantum correlation. It is quantum non-locality. 
Bell inequality \cite{Bell1969} is the first celebrated example of it by exhibiting a violation of its inequality. Further, the simpler version of it called Clauser, Horne, Shimony, and Holt (CHSH) inequality \cite{CHSH1969}, elucidating quantum non-locality between two spins the measurement outcome of which is dichotomic. For a singlet entangled state, the CHSH inequality explicitly breaks, describing a boundary line between quantum non-locality and local classical phenomena described by a hidden variable theory. Here we are confronted with two concepts of correlation, quantum entanglement and quantum non-locality. An important issue is to clarify how these physical concepts are related or different from each other. Entanglement may not be identical to the concept of quantum non-locality measured by Bell or CHSH inequality. As an amazing example, a mixed state with entanglement does not exhibit quantum non-locality \cite{Werner1989,Augusiak2015}, while entanglement and quantum non-locality can be identical in two qubit system \cite{Gisin1991}. Further study about this issue is required.

The study of quantum non-locality in many-body systems has been less-attracted than that of many-body entanglement. In particular, in a highly quantum correlated state, how quantum non-locality and further multipartite one appear and what structure of quantum non-locality many-body systems have are essentially important issues. 
The study of quantum non-locality can complement understanding the broad extensive study of quantum entanglement. 
Several studies have been reported along this context, a quantum phase transition in many-body system is characterized from the view of quantum non-locality in some complex lattice models \cite{Campbell2010,Justino2012,Huang2013,Sun2014,Lee2020,Wen2022,Liu2022} and disordered models \cite{Getelina2018,Liang2023}. But there is no unified understanding. Also, there is an interesting conjecture, two qubit quantum non-locality characterized by a violation of CHSH inequality does not occur in (discrete) translational invariant systems \cite{Oliveira2012}. 

This work contributes to the understanding of the appearance of quantum non-locality in many-body systems along the context discussed above. 
We study how non-locality appears in the internal structure in a non-interacting spinless fermion chain under quasi-periodic disorder, called extended Harper model\cite{Hatsugai1990,Han1994,Liu2015}. 
For fermion many-particle states of this model, two site and quantum non-locality is quantified by the CHSH inequality and further multi-site quantum non-locality, which is three site non-locality quantified by a Mermin–Klyshko-Svetlichny (MKS) polynomial \cite{Svetlichny1987,Mermin1990,Belinskii,Collins2002,Bancal2009}. 
We clarify the internal structure of such quantum non-locality estimated by the CHSH function and MKS polynomial, which has not been clarified so far. 
Especially, due to $U(1)$ conservation (particle number conservation) the CHSH and MKS polynomial can be efficiently calculated. We show how quantum non-locality explicitly characterize non-interacting many-body fermion states and its extended and critical phase transitions. In the globally averaged picture of the maximum value of quantum non-locality is not broken, but within the internal of the system the violation probabilities of the CHSH inequality and MKS polynomial can become sufficiently finite. 

The rest of this paper is organized as follows. 
In Sec.~II, we introduce the target fermion model and its basic properties. 
In Sec.~III, the CHSH function characterizing quantum non-locality between two sites and its practical calculation method for this model are explained and numerical study is shown. 
In Sec.~IV, we further show the practical scheme of the calculation of MKS polynomial applicable to this model and numerical study of adjacent three site (qubit or spin) MKS polynomial. 
Section V is devoted for conclusion.

\section{Model}
In this work, we consider the extended Harper model \cite{Hatsugai1990,Han1994}, the Hamiltonian of which is described by
\begin{eqnarray}
H_f=\sum^{L-1}_{j=0}\biggr[J_{j}(f^\dagger_{j}f_{j+1}+\mbox{h.c.})+h_j n_j\biggl],
\label{model}
\end{eqnarray}
where $f^\dagger_j$ and $f_j$ are fermion creation and annihilation operators, the number operator is $n_j\equiv f^\dagger_jf_j$ and $L$ is the system size. 
Here, the hopping coupling and on-site potential are quasi-periodic disorders, 
\begin{eqnarray}
J_{j}&=&1+\Delta_1 \sin(2\pi \beta j+\phi_0),\nonumber\\
h_j&=&\Delta_2 \cos(2\pi \beta j+\phi_0),\nonumber
\label{AApotential}
\end{eqnarray}
where $\beta=\frac{\sqrt{5}-1}{2}$ (golden ratio) and $\phi_0$ is a phase shift. Throughout this work, we consider half filling (the total particle number is $L/2$), open boundary condition and the couplings of the quasi-periodic disorders are positive, $\Delta_{1(2)}\geq 0$.

\begin{figure}[t]
\begin{center} 
\vspace{0.5cm}
\includegraphics[width=5.5cm]{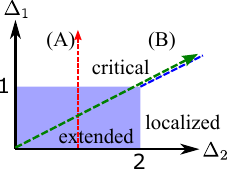}  
\end{center} 
\caption{Schematic phase diagram of the extended Harper model. There are three phases: extended, critical and localized phase. The phase boundary line between critical and localized phases is $2\Delta_1=\Delta_2$.}
\label{Fig1}
\end{figure}

So far, the phase diagram of the model of $H_{f}$ has been investigated in \cite{Han1994,Liu2015}. The schematic phase diagram is shown in Fig.~\ref{Fig1}. There are three phases: (i) Extended phase in the regime $\Delta_1\leq 1$ and $\Delta_2 \leq 2$, 
(ii) Critical phase in the regime $\Delta_1\geq 1$ and $\Delta_1\geq \frac{1}{2}\Delta_2$, (iii) Localized phase in the regime $\Delta_2\geq 2$ and $\Delta_1\leq \frac{1}{2}\Delta_2$. 
In the single particle picture, these states are true for all single particle eigenstates. There is no mobility edge in all phases \cite{Liu2015}.  
That is, if we consider excited states of many-body case the many-body states are in the same phase diagram. In particular, the critical phase includes broad correlation length \cite{Han1994}. Its explicit characterization of this phase is a subtle but important issue. In Ref.\cite{Liu2015}, a moderate value of participation ratio has been employed, which qualitatively identify the phase. We shall show that value of CHSH and MKS polynomials is a good indicator to identify the transition point.

We study the quantum non-locality of the many-body fermion state with $L/2$ fermion particles conserved, the state of which is described by
\begin{eqnarray}
|\Psi\rangle =\prod^{L/2-1}_{\ell=0}\psi^\dagger_{i_{\ell}}|0\rangle,
\label{state_def}
\end{eqnarray}
where $\psi^\dagger_{i_{\ell}}$ represents $i_{\ell}$-th single particle eigenstate of $H_{f}$ with $i_{\ell}$-th lowest single particle energy, described by $\psi^{\dagger}_{i_{\ell}}=\sum^{L-1}_{j=0}c^{i_\ell}_{j}f^\dagger_{j}$ where $\sum^{L-1}_{j=0}|c^{i_\ell}_{j}|^2=1$ and $|0\rangle$ is a vacuum state.
\section{Investigation of quantum non-locality by CHSH function}
Our interest is to investigate quantum non-local property in the non-interecting many-body feremion state in the model of $H_{f}$. To this end, we introduce the CHSH function, obtained from the correlations of two body spin operators. In the non-interacting fermion, such spin operator can be represented by Jordan-Wigner (JW) transformation \cite{LSM1961} shown in later.

\subsection{CHSH function}
The CHSH function is defined as 
\begin{eqnarray}
&&C(\vec{a}_1,\vec{a}_2,\vec{b}_1,\vec{b}_2)=\mathrm{tr}[\hat{\rho}_{\ell m}\hat{M}_{2}].
\label{CHSH_eq}
\end{eqnarray}
Here, 
\begin{eqnarray}
\hat{M}_2&=&\frac{1}{2}\biggr[\vec{a}_1\cdot \vec{\sigma}_\ell\otimes
(\vec{b}_1+\vec{b}_2)\cdot \vec{\sigma}_m\nonumber\\
&&+\vec{a}_2\cdot \vec{\sigma}_\ell\otimes
(\vec{b}_1-\vec{b}_2)\cdot \vec{\sigma}_m\biggl],
\end{eqnarray}
where $\ell,m$ represent a pair of spacial sites and $\vec{\sigma}_{\ell}$ is a Pauli operator vector described by $\vec{\sigma}_{\ell}=(\sigma^x_\ell,\sigma^y_{\ell},\sigma^z_{\ell})$, $\hat{\rho}_{\ell m}$ is a reduced density matrix on the two site pair $(\ell,m)$ and the set of normalized three-dimensional
vectors $(\vec{a}_1,\vec{a}_2,\vec{b}_1,\vec{b}_2)$ acts as optimization parameters. Then, $\hat{\rho}_{\ell m}$ is a reduced two site density matrix, given by a spin expectation value and spin-spin correlations \cite{Horodecki1995},
\begin{eqnarray}
\hat{\rho}_{\ell m}&=&\frac{1}{4}\biggl[I\otimes I
+(\vec{r}\cdot \vec{\sigma}_\ell)\otimes I 
+I \otimes (\vec{s}\cdot \vec{\sigma}_m)\nn
&+&\sum_{\alpha,\beta=x,y,z}t_{\alpha\beta}\sigma^\alpha_\ell\otimes \sigma^\beta_m\biggr],
\end{eqnarray}
where $I$ is a three dimensional identity matrix, $\vec{r}$, $\vec{s}$, and $t_{\alpha\beta}$ can be obtained by their expectation values of the many-body fermion state such as  
$\vec{r}=
(\langle \sigma^{x}_{\ell}\rangle,\langle \sigma^{y}_{\ell}\rangle, 
\langle \sigma^{z}_{\ell}\rangle
)
$,
$\vec{s}=
(\langle \sigma^{x}_{m}\rangle,\langle \sigma^{y}_{m}\rangle, 
\langle \sigma^{z}_{m}\rangle
)
$ and $t_{\alpha\beta}=\langle \sigma^{\alpha}_{\ell}\sigma^{\beta}_{m}\rangle$ 
since $\mathrm{tr}_{(\ell,m)}[\hat{\rho}_{\ell m}\hat{X}_{\ell m}]=\mathrm{tr}[\hat{\rho}_0\hat{X}_{\ell m}]$, in which $\hat{\rho}_0$ is a density matrix constructed by many-body state \cite{Cheong2004} and $\hat{X}_{\ell m}$ is an operator (observable) acting only on the sites, $\ell$ and $m$.

\subsection{Non-local violation regime of CHSH function}
As for a work \cite{Horodecki1995}, 
the maximum value of CHSH function given by Eq.~(\ref{CHSH_eq}) denoted by $C_{max}$ can be obtained without optimization for the set of the vectors $(\vec{a}_1,\vec{a}_2,\vec{b}_1,\vec{b}_2)$, 
\begin{eqnarray}
C_{max}=2\sqrt{\lambda_1+\lambda_2},
\label{CHSH_func}
\end{eqnarray}
where $\lambda_1$ and $\lambda_2$ are the largest and second largest eigenvalue of a $3\times 3$ matrix, $U=(T)^tT$ with $(T)_{\alpha,\beta}=t_{\alpha\beta}$. 
From the spin picture of the model of $H_{f}$, 
the spin-rotational symmetry in $x$-$y$ spin plane exists, then these correlation functions $t_{\alpha\beta}
$ satisfy 
$\langle \sigma^x_{\ell}\sigma^x_{m}\rangle=\langle \sigma^y_{\ell}\sigma^y_{m}\rangle\neq \langle \sigma^z_{\ell}\sigma^z_{m}\rangle$ \cite{come1}, and also $t_{\alpha\beta}=0$ for $\alpha \neq \beta$, straightforwardly obtained by JW fermion mapping and constraint of particle number conservation. Then, the matrix $U$ is diagonal. Thus, the maximum value of the CHSH is given by
\begin{eqnarray}
C_{max}=\max\{ 2\sqrt{2}|t_{xx}|,2\sqrt{|t_{zz}|^2+|t_{xx(yy)}|^2}\}.
\end{eqnarray}
Here, in quantum theory the maximal bound is known to be $C_{max}\leq 2\sqrt{2}$ \cite{Cirelson}. 
This is consistent to that the maximal correlation value $|t_{\alpha\beta}|\leq 1$. 
On the other hand, 
in the classical local theory (a hidden variable classical theory) \cite{CHSH1969}, 
the maximum value of $C(\vec{a}_1,\vec{a}_2,\vec{b}_1,\vec{b}_2)$ is bounded by $\max |C(\vec{a}_1,\vec{a}_2,\vec{b}_1,\vec{b}_2)|\leq 2$.
Thus, in this work, we regard the regime $2< C_{max}\leq 2\sqrt{2}$ as a quantum non-local violation regime. 
Our aim is to identify model-parameter regimes appearing the non-local violation regime by the CHSH function in the model of $H_f$.

\subsection{Calculation method of correlation functions in non-interacting many-body fermion states}
The CHSH function can be obtained by spin-spin correlation functions. 
According to the celebrated paper \cite{LSM1961}, the correlation functions can be straightforwardly treated in fermion picture from a set of single particle eigenstates, $\{(\epsilon_{k},\vec{\psi}^k)\}$ with $k=0,1,\cdots,L-1$, where $\vec{\psi}^k$ is $k$-th eigenstate of $H_f$, which is a real $L$-component column vector and $\epsilon_k$ $k$-th eigenenergy. The spin-spin correlation function can be obtained by employing the calculation method proposed by \cite{LSM1961,Sachdev1999,Osborne2002}. Here we make use of the particle conserving constraint. 

At first, we consider the $z$-component spin-spin correlation for a pair of site $(\ell,m)$. By using the JW transformation,  
\begin{eqnarray}
\sigma^x_j&=&f^\dagger_{j}e^{i\pi\sum^{j-1}_{k=0}n_k}+e^{-i\pi\sum^{j-1}_{k=0}n_k}f_j,\nonumber\\
\sigma^y_j&=&-i\biggr[f^\dagger_{j}e^{i\pi\sum^{j-1}_{k=0}n_k}+e^{-i\pi\sum^{j-1}_{k=0}n_k}f_j\biggl],\nonumber\\
\sigma^z_j&=&2n_j-1,
\end{eqnarray}
the correlation is described by 
\begin{eqnarray}
\langle \sigma^z_{\ell}\sigma^z_{m}\rangle&=&[G_{\ell,\ell}G_{m,m}-G_{m,\ell}G_{\ell, m}],
\end{eqnarray}
where $G_{m,m}$ is the lattice Green function \cite{Cheong2004,Peschel2009},   
\begin{eqnarray}
G_{i,j}\equiv 
\begin{cases}
    \langle f^\dagger_if_j\rangle+\langle f^\dagger_jf_i\rangle & \text{($i\neq j$)} \\
    2 \langle f^\dagger_if_j\rangle-1 & \text{($i=j$)}, 
  \end{cases}
\end{eqnarray}
and
\begin{eqnarray}
\langle f^\dagger_if_j\rangle&=&\sum_{q\in \Lambda_{L/2}}\vec{\psi}^{q}_{i}(\vec{\psi}^{q}_j)^\dagger.
\label{Corr_F}
\end{eqnarray}
$\Lambda_{L/2}$ represent a set of $L/2$ single particle eigenstates. 
Note that $G_{\ell,m}$ is symmetric, $G_{\ell,m}=G_{m,\ell}$ since the eigenvector $\vec{\psi}^{k}$ is real for any $k$.

Other spin components of correlations are similarly given by 
\begin{eqnarray}
\langle \sigma^x_{\ell}\sigma^x_{m}\rangle&=&\mathrm{det}\tilde{G}(\ell,m),\\
\tilde{G}(\ell,m)&\equiv&
\begin{pmatrix} 
  G_{\ell,\ell+1} & G_{\ell,\ell+2} & \dots  & G_{\ell,m} \\
  G_{\ell+1,\ell+1} & G_{\ell+1,\ell+2} & \dots  & G_{\ell+1,m} \\
  \vdots & \vdots & \ddots & \vdots \\
  G_{m-1,\ell+1} & G_{m-1,\ell+2} & \dots  & G_{m-1,m}
\end{pmatrix},\nonumber
\end{eqnarray}
and
\begin{eqnarray}
\langle \sigma^y_{\ell}\sigma^y_{m}\rangle&=&
\mathrm{det}\bar{G}(\ell,m),\\
\bar{G}(\ell,m)&\equiv&
\begin{pmatrix} 
  G_{\ell+1,\ell} & G_{\ell+1,\ell+1} & \dots  & G_{\ell,m-1} \\
  G_{\ell+2,\ell} & G_{\ell+2,\ell+1} & \dots  & G_{\ell+1,m-1} \\
  \vdots & \vdots & \ddots & \vdots \\
  G_{m,\ell} & G_{m,\ell+1} & \dots  & G_{m,m-1}
\end{pmatrix}.\nonumber
\end{eqnarray}

Note that since the model of $H_s$ is not translational invariant, the above spin-spin correlation functions are not described only by the distance of the pair of sites $r=|\ell-m|$.   

\subsection{Numerical observation for CHSH function}
We numerically observe the detailed behavior of the CHSH function, and treat many different samples with different uniform random phase shift $\phi_0$. Each sample is labeled by $s$ and the total number of samples is denoted by $N_s$. 

In this numerical study, the following quantities are considered. 
The first is an averaged value of the maximum value of CHSH defined as
\begin{eqnarray}
\langle C_{max}\rangle (r)\equiv \frac{1}{N_s}\sum_{s}\biggl[\frac{1}{L-r}\sum^{L-r-1}_{j=0}C^{s}_{max}(j,j+r)\biggr].
\label{CMAX}
\end{eqnarray}
Here $C^{s}_{max}(j,j+r)$ is the value of CHSH function of Eq.~(\ref{CHSH_func}) of a sample $s$, where the pair of site is $j$ and $j+r$, $r$ is a distance between the pair. 
From the above expression, the value of $\langle C_{max}\rangle (r)$ represents a global property of the degree of non-locality in the entire system.

As the second quantity, we introduce the probability of the local CHSH violation event denoted by $p_v$. 
This quantity is defined as
\begin{eqnarray}
p_v(r)&\equiv&\frac{1}{N_s}\sum_{s}v_s(r),\\
v_s(r)&\equiv&\frac{1}{L-r}\sum^{L-r-1}_{j=0}H_0(C^{s}_{max}(j,j+r)-2).
\end{eqnarray}
Here $H_0(x)$ is the Heaviside step function where $H_0(x)=1$ for $x>0$, otherwise, $H_0(x)=0$. That is, if the CHSH function a pair of sites in the system is in the non-local violation regime, the event is counted by $H_0(x)$. 
The quantity $p_v(r)$ evaluates whether CHSH inequality breaking is occurring somewhere in a two-site pair within the system. 
It gives information on the CHSH inequality breaking from the local point view, giving a different point of view from the globally average value of the CHSH function of Eq.~(\ref{CMAX}).

In the numerical calculation, two different many-body states are considered: 
(i) $L/2$ particle ground state, corresponding to the set $\{i_\ell\}=\{i_0,i_1,\cdots, i_{L/2-1}\}=\{0,1,2,\cdots, L/2-1\}$ in the state definition of Eq.~(\ref{state_def}), denoted as $|\psi_g\rangle$, (ii) an typical excited state, set by $\{i_\ell\}=\{i_0,i_1,\cdots, i_{L/2-1}\}=\{L/4,L/4+1,L/4+2,\cdots, 3L/4-1\}$ in the state definition of Eq.~(\ref{state_def}), denoted as $|\psi_e\rangle$ and 
we focus on two parameter sweeps as shown in Fig.~\ref{Fig1}: (A) varying $\Delta_1$ with $\Delta_2=1$ corresponding to the parameter sweep between the extended and critical phases, 
(B) varying $\Delta_1$ with the condition $\Delta_2= 2\Delta_1$, corresponding to the parameter sweep between the extended and phase boundary line of the critical and localized phases. 

\begin{figure}[t]
\begin{center} 
\vspace{0.5cm}
\includegraphics[width=8.5cm]{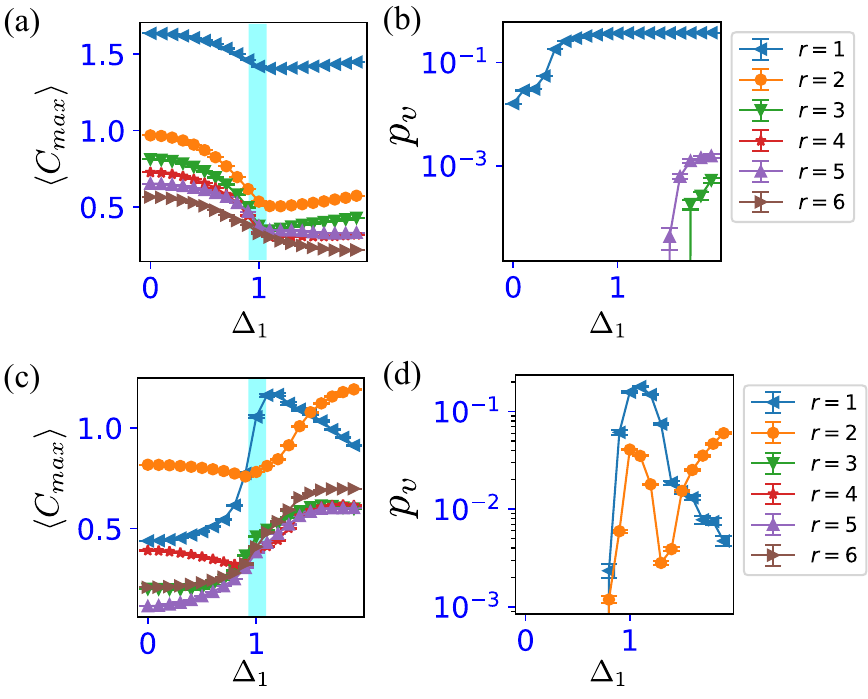}  
\end{center} 
\caption{ 
(a) Average value $\langle C_{max}\rangle$ for the groundstate $|\psi_g\rangle$. (b) Internal violation probability $p_v$ for the groundstate $|\psi_g\rangle$. 
(c) Average value $\langle C_{max}\rangle$ for the excited state $|\psi_e\rangle$. (d) Internal violation probability $p_v$ for the excited state $|\psi_e\rangle$.
For all data, the parameter sweep is of (A), where $\Delta_2=1$, and the distance of two site pair is chosen by $r=1-6$. The system size is $L=64$ and the number of $\phi_0$ samples is $7 \times 10^3$.}
\label{Fig2}
\end{figure}

Figure \ref{Fig2} shows the numerical results for the average value $\langle C_{max}\rangle$ and the internal violation probability $p_v$ for the groundstate $|\psi_g\rangle$ and the excited state $|\psi_e\rangle$, where we set the parameter sweep (A) and observe the distance of two site pair up to $r=6$.

See Fig.~\ref{Fig2}(a), we observe that $\langle C_{max}\rangle$ exhibits a drop around $\Delta_1=1$ for any $r$ for the state $|\psi_g\rangle$, indicating the presence of the phase transition between the extended and critical phases, consistent to the previous works \cite{Han1994,Liu2015}. That is, $\langle C_{max}\rangle$ is a good indicator to identify a phase transition, and we also find that the entire value of $\langle C_{max}\rangle$ on the parameter sweep does not break the CHSH inequality, any non-local violation does not appear globally in the system for any distance $r$, both phases and transition points and $r=1$ value of $\langle C_{max}\rangle$ is larger than any other $r$ case.  
We turn to observe $p_v$ as Fig.~\ref{Fig2}(b). 
Interestingly, a finite value of $p_v$ ($\mathcal{O}(10^{-1})$) appears for $r=1$ case from the regime near to critical phase but not for $r>1$. This implies that the CHSH inequality breaks down {\it with a finite probability in a portion of the internal of the system}. We also note that
the violation probability $p_v$ in the critical phase is larger than that of extended phase but $\langle C_{max}\rangle$ in the critical phase is smaller than that of extended phase. In addition, in Fig.~\ref{Fig2}(b), $p_v$ exhibits a much small finite value for $r=3,5$ (generally, $r$ is in the odd case). The data implies the non-local violation may occur as a rare event in the deep critical phase.

We next focus on the case of the excited state $|\psi_e\rangle$ for the parameter sweep (A).
In Fig.~\ref{Fig2}(c), $\langle C_{max}\rangle$ exhibits sudden increase around $\Delta_1=1$ for any $r$, indicating the presence of the phase transition of the excited state $|\psi_e\rangle$ between the extended and critical phases. The entire value of $\langle C_{max}\rangle$ on the parameter sweep indicates that  the CHSH inequality does not break down globally. However, we turn to observe $p_v$ as Fig.~\ref{Fig2}(d). Interestingly, we find that a peak of a finite value of $p_v$ ($\mathcal{O}(10^{-1})$) appears for $r=1$ and also for $r=2$ cases a finite value ($\mathcal{O}(10^{-2})$) appears. This implies that the non-local quantum  correlation for $r=1$ and $2$ occurs as a breakdown of the CHSH inequality with a finite probability in a portion of the internal system around the phase transition. We also note that
the violation probability $p_v$ for $r=2$ once exhibit a sudden increase around $\Delta_1=1$ and then decreases suddenly after passing $\Delta_1=1$ and increase again with increase $\Delta_1$. This peculiar behavior 
suggests that the phase change between the extended and critical phases may not be a single phase transition. A similar behavior appears in the later calculation of the MKS polynomial. 

\begin{figure}[t]
\begin{center} 
\vspace{0.5cm}
\includegraphics[width=8.5cm]{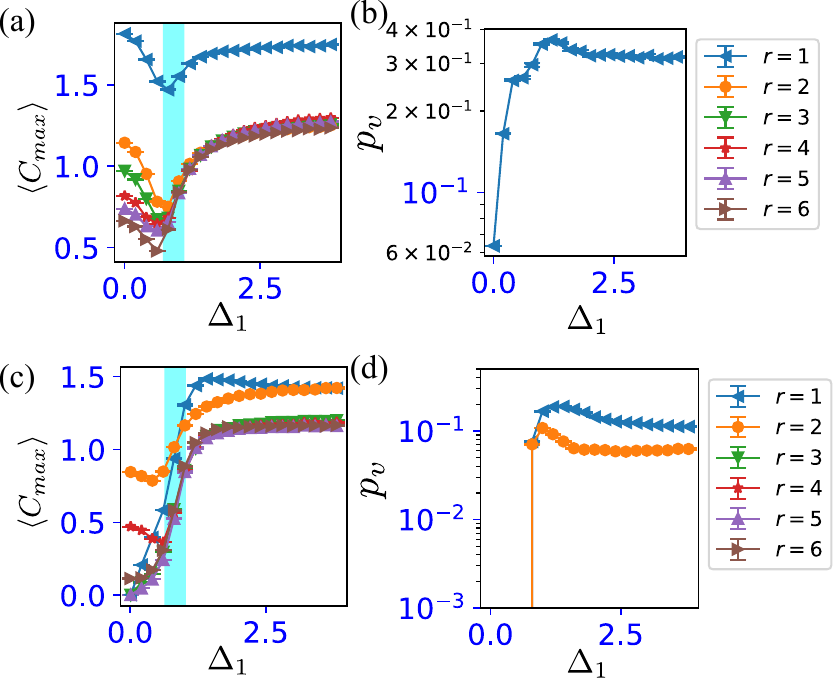}  
\end{center} 
\caption{(a) Average value $\langle C_{max}\rangle$ for the groundstate $|\psi_g\rangle$. (b) Internal violation probability $p_v$ for the groundstate $|\psi_g\rangle$. 
(c) Average value $\langle C_{max}\rangle$ for the excited state $|\psi_e\rangle$. (d) Internal violation probability $p_v$ for the excited state $|\psi_e\rangle$.
For all data, the parameter sweep is of (B) and the distance of two site pair is chosen by $r=1-6$. The system size is $L=64$ and the number of $\phi_0$ samples is $7 \times 10^3$.}
\label{Fig3}
\end{figure}

We move to the case of the parameter sweep (B). The behavior of $\langle C_{max}\rangle$ for the groundstate $|\psi_g\rangle$ is displayed in Fig.~\ref{Fig3}(a). 
All data exhibits a drop around $\Delta_1=1$ for any $r$ for the state $|\psi_g\rangle$, indicating the presence of the phase change of the ground state between the extended phase and the regime on the phase boundary between the critical and localized phases, again consistent to the previous works \cite{Han1994,Liu2015}. 
$\langle C_{max}\rangle$ well characterize the vanishing extended phase. The entire value of $\langle C_{max}\rangle$ on the parameter sweep indicates that  the CHSH inequality does not break down globally in the system for any distance $r$. Also, the value of $\langle C_{max}\rangle$ for $r\geq 2$ is same on the boundary, implying a strong criticality, that is, the degree of correlation exhibits scale free tendency. 
We further observe $p_v$ for the groundstate $|\psi_g\rangle$ as Fig.~\ref{Fig3}(b). 
Interestingly, a large finite value of $p_v$ ($\mathcal{O}(10^{-1})$) appears only for $r=1$ case but not for $r>1$ and the regime starts around a regime vanishing the extended phase $\Delta_1\sim 1$. Only the nearest neighbor correlation exhibits quantum non-locality as the break down of the CHSH inequality with a finite probability in a portion of the internal of the system on the phase boundary between the critical and localized phases.

We further observe the case of the excited state $|\psi_e\rangle$. 
In Fig.~\ref{Fig3}(c), $\langle C_{max}\rangle$ exhibits a sudden increase around $\Delta_1=1$ for any $r$, indicating the presence of vanishing the extended phase in the excited state $|\psi_e\rangle$. The entire value of $\langle C_{max}\rangle$ on the parameter sweep indicates that the CHSH inequality does not break down globally in the system for any distance $r$. However, we find that as shown in Fig.~\ref{Fig3}(d), a finite value of $p_v$ ($\mathcal{O}(10^{-1})$) appears for $r=1$ and $r=2$ cases after stating to vanish the extended phase. This implies that the correlation for $r=1$ and $2$ exhibits quantum nonlocality as the break down of the CHSH inequality with a finite probability in a portion of the internal system on the phase boundary between the critical and localized phases. 

We also care the system size dependence of $\langle C_{max}\rangle$ and $p_v$. The properties of the transition and its entire behavior are almost same. The results are shown in Appendix C.

\section{Investigation of three-qubit MKS polynomial}

We next observe an extension form of the CHSH function. 
The general form is called Mermin–Klyshko-Svetlichny (MKS) polynomial \cite{Svetlichny1987,Mermin1990,Belinskii,Belinskii,Collins2002}, which can characterize $n$-qubit multipartite quantum non-locality, also can capture its hierarchical structure of non-locality \cite{Wen2022}. The brief explanation of general form of the MKS polynomial is in Appendix A.  

In this numerical study, we limit adjacent three-qubit MKS (3MKS) polynomial.  
The concrete form for a three-qubit subsystem (three spatial sites) we consider here is given as the expectation value of 3MKS polynomial \cite{Svetlichny1987,Mermin1990,Belinskii,Collins2002}
\begin{eqnarray}
&&C^{3MKS}({\bf a})=\frac{1}{2}\biggr[\mathrm{tr}_{3}(\hat{\rho}_3 \hat{M}_3)+\mathrm{tr}_{3}(\hat{\rho}_3 \hat{M}'_3)\biggl],
\label{3MKS}
\end{eqnarray}
where ${\bf a}=(\vec{a}_{j_1},\vec{a}'_{j_1},\vec{a}_{j_2},\vec{a}'_{j_2},\vec{a}_{j_3},\vec{a}'_{j_3})$, corresponds to the set of optimization vector parameters, 
$\hat{\rho}_3$ is an adjacent three-site reduced density matrix and $\hat{M}^{(')}_3$ is 3MKS operator, constituted by some lower-lank operators as
\begin{eqnarray}
&&\hat{M}_3=\frac{1}{2}\hat{M}_2
\otimes (\vec{a}_{j_3}+\vec{a}'_{j_3})\cdot \vec{\sigma}_{j_3}
+
\frac{1}{2}\hat{M}'_2
\otimes (\vec{a}_{j_3}-\vec{a}'_{j_3})\cdot \vec{\sigma}_{j_3},\nonumber\\
&&\hat{M}'_3=\frac{1}{2}\hat{M}'_2
\otimes (\vec{a}_{j_3}+\vec{a}'_{j_3})\cdot \vec{\sigma}_{j_3}
+
\frac{1}{2}\hat{M}_2
\otimes (\vec{a}'_{j_3}-\vec{a}_{j_3})\cdot \vec{\sigma}_{j_3},\nonumber\\
&&\hat{M}_2=\frac{1}{2}\hat{M}_1
\otimes (\vec{a}_{j_2}+\vec{a}'_{j_2})\cdot \vec{\sigma}_{j_2}
+
\frac{1}{2}\hat{M}'_1
\otimes (\vec{a}_{j_2}-\vec{a}'_{j_2})\cdot \vec{\sigma}_{j_2},\nonumber\\
&&\hat{M}'_2=\frac{1}{2}\hat{M}'_1
\otimes (\vec{a}_{j_2}+\vec{a}'_{j_2})\cdot \vec{\sigma}_{j_2}
+
\frac{1}{2}\hat{M}_1
\otimes (\vec{a}'_{j_2}-\vec{a}_{j_2})\cdot \vec{\sigma}_{j_2},\nonumber\\
&&\hat{M}_1=\vec{a}_{j_1}\cdot \vec{\sigma}_{j_1},\:\:
\hat{M}'_1=\vec{a}'_{j_1}\cdot \vec{\sigma}_{j_1},\nonumber
\end{eqnarray}
where each optimization vector $\vec{a}^{(')}_j$ is a spherical unit vector parameterized by two angle parameters as $\vec{a}^{(')}_j=(\cos\theta^{(')}_{j} \sin\phi^{(')}_{j},
\sin\theta^{(')}_{j} \sin\phi^{(')}_{j},
\cos\phi^{(')}_{j})$ and the label $(j_1,j_2,j_3)$ represents a three adjacent lattice sites in the system. 
Our interest is to estimate a maximum value of $C^{3MKS}$ denoted by $C^{3MKS}_{max}$. To this end, we solve the optimization problem for 12 rotational angle parameters by employing a numerical Scipy package \cite{Scipy_opt}.

In the local classical theory (a local hidden variable theory), the maximum value of $C^{3MKS}_{max}$ is bounded by $C^{3MKS}_{max}\leq 1$ \cite{Gisin1998,Collins2002}. On the other side, in quantum theory, the maximum value of $C^{3MKS}_{max}$ is bounded by $C^{3MKS}_{max}\leq \sqrt{2}$, where $C^{3MKS}_{max}=\sqrt{2}$ indicates full three-body non-locality among the three site. 
Thus, the quantum non-locality for the three-sites is characterized by $1<C^{3MKS}_{max} \leq \sqrt{2}$, called 3MKS non-local violation regime.

\begin{figure}[t]
\begin{center} 
\vspace{0.5cm}
\includegraphics[width=8.5cm]{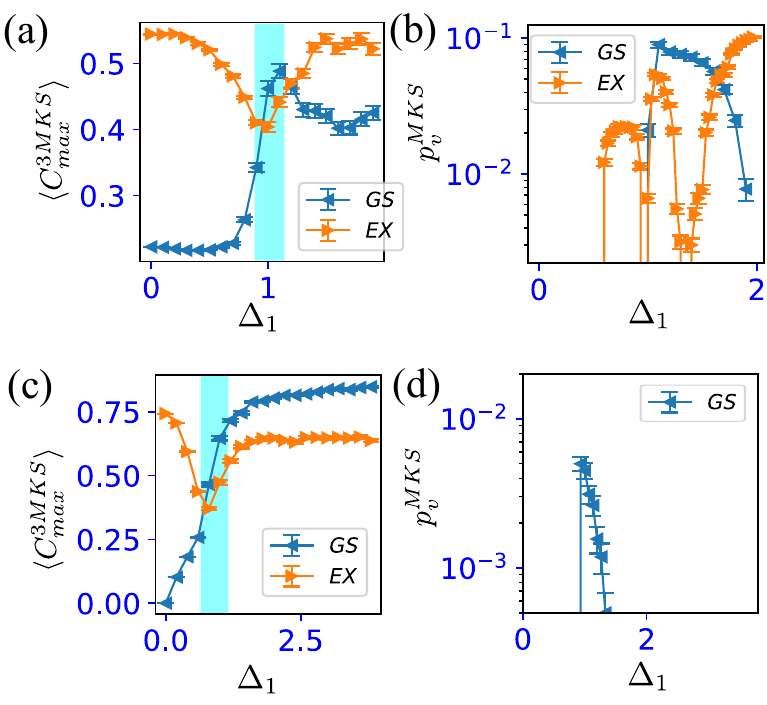}  
\end{center} 
\caption{(a) Average value $\langle C^{3MKS}_{max}\rangle$ for the parameter sweep (A). 
(b) Internal violation probability $p^{MKS}_v$ for the parameter sweep (A).  
(c) Average value $\langle C^{3MKS}_{max}\rangle$ for the parameter sweep (B). 
(d) Internal violation probability $p^{MKS}_v$ for the parameter sweep (B). 
The system size is $L=64$ and the number of $\phi_0$ samples is $8 \times 10^3$. The labels ``GS" and ``EX" represent the states $|\psi_g\rangle$ and $|\psi_e\rangle$.}
\label{Fig4}
\end{figure}

As an example, in only three qubit system, the GHZ state $\displaystyle{|{\rm GHZ}\rangle =\frac{1}{\sqrt{2}}[|111\rangle+|000\rangle]}$ is known to give a maximum quantum non-locality, $C^{3MKS}_{max}=\sqrt{2}$ \cite{Collins2002}. 

Let us move to numerical observations about the detailed behavior of the 3MKS, as same to the case of the CHSH function. We set the three adjacent sites around the center of the system, $j_1=L/2-1$, $j_2=L/2$, and $j_3=L/2+1$, and treat many different samples with different $\phi_0$. 
We explain the detail of the calculation of 3MKS for the fermion many-body system as shown in Appendix B. 
We observe the averaged value of the maximum value of 3MKS, 
\begin{eqnarray}
\langle C^{3MKS}_{max}\rangle\equiv \frac{1}{N_s}\sum_{s}C^{3MKS,s}_{max},
\end{eqnarray}
where we denote 3MKS of a sample $s$ by $C^{3MKS,s}_{max}$, 
and also observe the violation event probability 
\begin{eqnarray}
p^{MKS}_v&\equiv&\frac{1}{N_s}\sum_{s}v^{MKS}_s,
\end{eqnarray}
with $v^{MKS}_s\equiv H_0(C^{3MKS,s}_{max}-1)$. 
As same to the case of the calculation of the CHSH function, we pick up the two parameter sweeps (A) and (B), and focus on the two many-body states, $|\psi_g\rangle$ and $|\psi_{e}\rangle$. 

The behavior of $\langle C^{3MKS}_{max}\rangle$ for the parameter sweep (A) is displayed in Fig.~\ref{Fig4}(a). The data of $|\psi_g\rangle$ exhibits a peak at $\Delta_1=1$ and the data of $|\psi_e\rangle$ exhibits a drop at $\Delta_1=1$, indicating the presence of the phase transition in the different behavior. $\langle C^{3MKS}_{max}\rangle$ well characterize the phase transition. The entire value of $\langle C^{3MKS}_{max}\rangle$ on the parameter sweep does not enter into the 3MKS non-local violation regime. 
But, we further observe $p^{3MKS}_v$ for the parameter sweep (A) as shown in Fig.~\ref{Fig4}(b). 
We find that for the state $|\psi_g\rangle$, 
a sudden increase of $p^{3MKS}_v$ up to $\mathcal{O}(10^{-1})$ appears around $\Delta_1=1$ and the 3MKS exhibits the non-local violation regime with a finite probability in the internal of the system. 
We move to observe the data for the state $|\psi_e\rangle$. 
We find that for the state $|\psi_e\rangle$, 
$p^{3MKS}_v$ starts to have a finite value up to $\mathcal{O}(10^{-1})$ from entering the critical phase  and the 3MKS exhibits the non-local violation regime with a finite probability in the internal of the system.
Further, note that a large fluctuation behavior around $\Delta_1=1$ is observed. This behavior is similar to the behavior of the CHSH function shown in $r=2$ data of Fig.~\ref{Fig2}(d), indicating 
the phase change between the extended and critical phases may not be a single phase transition.

Finally, we turn to the results for the parameter sweep (B). 
In Fig.~\ref{Fig4}(c), $\langle C_{max}\rangle$ exhibits similar behavior to the case of the parameter sweep (A) and characterize vanishing the extended phase for both states $|\psi_g\rangle$ and $|\psi_e\rangle$. The entire value of $\langle C_{max}\rangle$ on the parameter sweep does not exhibit the MKS non-local violation regime. 
As shown in Fig.~\ref{Fig4}(d), the event of the MKS non-local violation occurs only around the phase change as observed that only for the state $|\psi_g\rangle$, a finite value of $p_v$ ($\mathcal{O}(10^{-2})$) appears at $\Delta_1=1$. This indicate that the event of the MKS non-local violation occurs around the tricritical point of the model in the ground state $|\psi_g\rangle$ with a finite probability.

\section{Conclusion}
We showed that the CHSH function and MKS polynomial are good indicators to characterize the phase transitions in disordered fermion systems. As an example, we studied non-local properties in the extended Harper model. 
We found that the non-local violation regime of both the CHSH function and 3MKS locally appears in a critical phase and around its transition points with a finite probability. Such a violation does not appear as an averaged value. 
Further, the CHSH function and 3MKS exhibits similar behavior and they are also useful for detecting a phase transition for excited states. These quantities can be useful for understanding the detailed structure of some complex phase transitions in disordered systems.

We expect that this characterization approach suggested in this work has broad applications for various non-interacting systems, such as long-range hopping model, e.g. how the long-range nature changes properties of quantum non-locality in many-body systems. 

\bigskip
\section*{ACKNOWLEDGMENTS}
The work is supported by JSPS
KAKEN-HI Grant Number 23K13026 (Y.K.), Japan.

\appendix
\renewcommand{\theequation}{A.\arabic{equation} }
\setcounter{equation}{0}

\section*{Appendix}

\section*{A. General MKS polynomial}
General MKS polynomials for $n$-qubits can be recursively defined depending on even or odd $n$ \cite{Collins2002}. 
The two types of MKS polynomias are introduced as the following forms
\begin{eqnarray}
S^{MKS,1}_{n}({\bf a})&=&\mathrm{tr}(\hat{\rho}_n \hat{M}_n),\\ 
S^{MKS,2}_{n}({\bf a})&=&\mathrm{tr}(\hat{\rho}_n \hat{M}^{'}_n).
\label{2pre_n-MKS_}
\end{eqnarray}
Here $\hat{\rho}_n$ is a reduced density matrix for $n$-qubit subsystem in a system, ${\bf a}=(\vec{a}_{j_1},\vec{a}'_{j_1},\vec{a}_{j_2},\vec{a}'_{j_2},\cdots,\vec{a}_{j_n},\vec{a}'_{j_n})$ including total $4n$ optimization parameters, and the above $n$-qubit MKS operator $\hat{M}_n$ is recursively defined by \cite{Gisin1998}
\begin{eqnarray}
\hat{M}_n&=&\frac{1}{2}\hat{M}_{n-1}\otimes (\vec{a}_{j_n}
+\vec{a}^{'}_{j_n})\cdot {\vec{\sigma}}_{j_n}\nonumber\\
&&+\frac{1}{2}\hat{M}^{'}_{n-1}\otimes (\vec{a}_{j_n}
-\vec{a}^{'}_{j_n})\cdot {\vec{\sigma}}_{j_n},
\label{n-MKS}
\end{eqnarray}
where the operator $\hat{M}^{'}_n$ is obtained from $\hat{M}_n$ by exchanging all the
primed and non-primed $\vec{a}$’s and 
$n=1$ case are given by $\hat{M}_1=\vec{a}_{j_1}\cdot \vec{\sigma}_{j_1}$ and $
\hat{M}'_1=\vec{a}'_{j_1}\cdot \vec{\sigma}_{j_1}$.

As for Ref.\cite{Collins2002}, 
as a measure of quantum non-locality, the form of the generalized MKS polynomial depends on whether $n$ is even or odd as
\begin{eqnarray}
C^{nMKS}\equiv 
\begin{cases}
    S^{MKS,1}_{n}({\bf a}) & \text{($n=$even)} \\
    \displaystyle{\frac{1}{2}\biggr[S^{MKS,1}_{n}({\bf a})+S^{MKS,2}_{n}({\bf a})\biggl]} & \text{($n=$odd)}
  \end{cases}.\nonumber\\
\label{2pre_n-MKS_}
\end{eqnarray}
The maximum optimized value denoted by $C^{nMKS}_{max}$ ($\geq 0$) of the above can evaluates $n$-qubit (or $n$-site) non-locality. If the theory is local and classical (a local hidden variable model),  $C^{nMKS}_{max}\leq 1$. On the other hand, if the theory exhibits a quantum non-locality, the above quantity $C^{nMKS}_{max}$ is in regimes; 
$1< C^{nMKS}_{max} \leq \sqrt{2^{n-1}}$ for even $n$ and $1< C^{nMKS}_{max} \leq \sqrt{2^{n-2}}$ for odd $n$. That is, the maximum bound values for the quantum non-locality are
$\sqrt{2^{n-1}}$ for even $n$ and $\sqrt{2^{n-2}}$, respectively. 

In this work, for $n=2$ case, we observe $C_{max}\equiv 2C^{nMKS}_{max}$ by multiplying a factor $2$ respecting the form of CHSH inequality \cite{CHSH1969}, while we use the form of $C^{nMKS}_{max}$ of the MKS polynomial for $n=3$ case.

\section*{B. Detail of calculation of $n=3$ MKS polynomial from fermion picture}

The 3MKS of Eq.~(\ref{3MKS}) is represented by a sum of three-spin correlation functions,
\begin{eqnarray}
C^{3MKS}
&=&\frac{1}{2}\biggl[\mathrm{tr}(\hat{\rho}_3\hat{M}_2
\otimes (\vec{a}^{'}_{j_3}\cdot {\vec{\sigma}}_{j_3}))\nonumber\\
&&+\mathrm{tr}(\hat{\rho}_3\hat{M}'_2\otimes(\vec{a}_{j_3}\cdot  \vec{\sigma}_{j_3}))\biggr]\nonumber\\
&=&\frac{1}{2}\biggl[\mathrm{tr}(\hat{\rho}_3\hat{M}_2
\otimes (\vec{a}^{'}_{j_3}\cdot {\vec{\sigma}}_{j_3}))\nonumber\\
&&+\mathrm{tr}(\hat{\rho}_3\hat{M}'_2\otimes(\vec{a}_{j_3}\cdot  \vec{\sigma}_{j_3}))\biggr]\nonumber\\
&=&\sum_{(\alpha,\beta,\gamma)\in \mathcal{S}}
\frac{1}{2}\biggr[a^{\alpha}_{j_1} a^{\beta}_{j_2} a'^{\gamma}_{j_3}
+a'^{\alpha}_{j_1}a^{\beta}_{j_2} a'^{\gamma}_{j_3}\nonumber\\
&&+a^\alpha_{j_1} a'^\beta_{j_2} a'^\gamma_{j_3}
-a'^\alpha_{j_1} a'^\beta_{j_2} a'^\gamma_{j_3}\biggl]
\langle \sigma^{\alpha}_{j_1}\sigma^{\beta}_{j_2}\sigma^\gamma_{j_3}\rangle\nonumber\\
&&+\sum_{(\alpha,\beta,\gamma)\in \mathcal{S}}
\frac{1}{2}\biggr[a'^{\alpha}_{j_1} a'^{\beta}_{j_2} a^{\gamma}_{j_3}
+a^{\alpha}_{j_1}a'^{\beta}_{j_2} a^{\gamma}_{j_3}\nonumber\\
&&+a'^\alpha_{j_1} a^\beta_{j_2} a^\gamma_{j_3}
-a^\alpha_{j_1} a^\beta_{j_2} a^\gamma_{j_3}\biggl]
\langle \sigma^{\alpha}_{j_1}\sigma^{\beta}_{j_2}\sigma^\gamma_{j_3}\rangle,\nonumber\\
\end{eqnarray}
where $\langle \cdot \rangle$ means expectation value for the many-body state $|\Psi\rangle$ of Eq.~(\ref{state_def}), the optimization parameters are
\begin{eqnarray}
\vec{a}_{j}(\theta_j,\phi_j)&\equiv& (a^x_{j},a^y_{j},a^z_{j})\nonumber\\
&=&(\cos\theta_{j}\sin\phi_{j},\sin\theta_{j}\sin\phi_{j},\cos\phi_{j}),\nonumber\\
\vec{a}'_{j}(\theta'_j,\phi'_j)&\equiv& (a'^x_{j},a'^y_{j},a'^z_{j})\nonumber\\
&=&(\cos\theta'_{j}\sin\phi'_{j},\sin\theta'_{j}\sin\phi'_{j},\cos\phi'_{j})\nonumber,
\end{eqnarray}
and 
\begin{eqnarray}
\mathcal{S}=&&\{(z,z,z),(x,z,x),(z,x,x),\nonumber\\
&&(x,x,z),(y,z,y),(y,y,z),(z,y,y)\}.\nonumber
\end{eqnarray} 
The above sums over the labels of spin component, $(\alpha,\beta,\gamma)$. 
The restricted choice of $\mathcal{S}$ originates from the constraint of particle number conservation in the fermion model. 

Practically, each expectation values of  $\langle\sigma^{\alpha}_{j_1}\sigma^{\beta}_{j_2}\sigma^\gamma_{j_3}\rangle$  with $(\alpha,\beta,\gamma)\in \mathcal{S}$ can be calculated by transforming them into the fermion picture though the Jordan-Winger transformation. 
The concrete forms for correlations given by the element of $\mathcal{S}$ are given as follows:
\begin{eqnarray}
\langle \sigma^x_{j_1}\sigma^z_{j_2}\sigma^x_{j_3}\rangle&=&
\langle f^\dagger_{j_1} e^{i\sum^{j_1-1}_{m=0}n_m}(2n_{j_2}-1)e^{-i\sum^{j_3-1}_{m=0}n_m}f_{j_3}\rangle\nonumber\\
&+&\langle (f_{j_1} e^{-i\sum^{j_1-1}_{m=0}n_m}(2n_{j_2}-1)e^{i\sum^{j_3-1}_{m=0}n_m}f^\dagger_{j_3})\rangle,
\nonumber\\
\langle \sigma^x_{j_1}\sigma^x_{j_2}\sigma^z_{j_3}\rangle&=&
\langle f^\dagger_{j_1}f_{j_2}(2n_{j_3}-1)\rangle
-\langle f_{j_1}f^\dagger_{j_2}(2n_{j_3}-1)\rangle,\nonumber\\
\langle \sigma^z_{j_1}\sigma^x_{j_2}\sigma^x_{j_3}\rangle&=&
\langle (2n_{j_1}-1)f^\dagger_{j_2}f_{j_3}\rangle
-\langle (2n_{j_1}-1)f_{j_2}f^\dagger_{j_3}\rangle,
\nonumber\\
\langle \sigma^z_{j_1}\sigma^z_{j_2}\sigma^z_{j_3}\rangle&=&
8\langle n_{j_1}n_{j_2}n_{j_3} \rangle\nonumber\\
&&-4\langle n_{j_1}n_{j_2}\rangle 
-4\langle n_{j_2}n_{j_3}\rangle 
-4\langle n_{j_3}n_{j_1}\rangle\nonumber\\
&&+2\langle n_{j_1}\rangle
+2\langle n_{j_2}\rangle
+2\langle n_{j_3}\rangle-1.\nonumber
\end{eqnarray}
By using Wick's theorem, each correlations can be further written as a multiple form of the fermion correlation functions of a non-interecting many-body state, $\langle f^\dagger_if_j\rangle$ obtained by Eq.~(\ref{Corr_F}) in the main text. 
Also, since the Hamiltonian of Eq.~(\ref{model}) has a symmetry of replacement $x\longleftrightarrow y$ for the spin components, some relationship like $\langle \sigma^x_{j_1}\sigma^z_{j_2}\sigma^x_{j_3}\rangle=\langle \sigma^y_{j_1}\sigma^z_{j_2}\sigma^y_{j_3}\rangle$ exists.
Also, this fact is straightforwardly checked only by representing them in fermion picture under particle conservation. 

\begin{figure}[b]
\begin{center} 
\vspace{0.5cm}
\includegraphics[width=8.5cm]{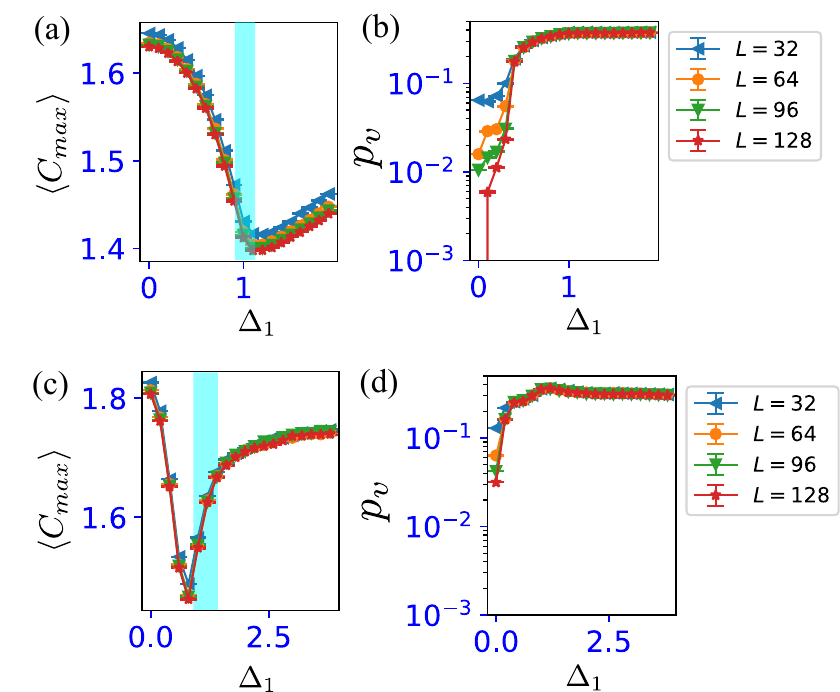}  
\end{center} 
\caption{System size dependence for average value $\langle C_{max}\rangle$ for $r=1$ [(a)] and internal violation probability $p_v$ for $r=1$ [(b)] in the parameter sweep (A).
System size dependence for average value $\langle C_{max}\rangle$ for $r=1$ [(c)] and internal violation probability $p_v$ for $r=1$ in the parameter sweep (B).
For all data, the number of $\phi_0$ samples is $3 \times 10^3$-$1\times 10^4$.}
\label{Sys_d}
\end{figure}

\begin{figure}[t]
\begin{center} 
\vspace{0.5cm}
\includegraphics[width=8.8cm]{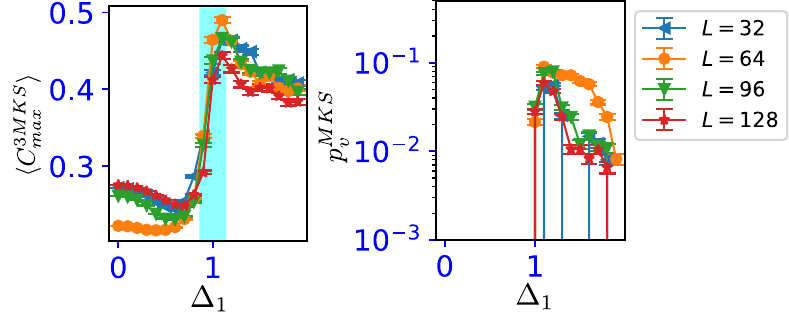}  
\end{center} 
\caption{System size dependence for average value $\langle C^{3MKS}_{max}\rangle$ [(a)] and internal violation probability $p^{MKS}_v$[(b)] in the parameter sweep (A). 
For all data, the number of $\phi_0$ samples is $7 \times 10^3$.}
\label{Sys_dd}
\end{figure}
As the result, if we consider the adjacent three site case $j_1=j_2-1=j_3-2$, then in this system we only need to calculate total seven patterns of the three body spin correlations coming from the set of $\mathcal{S}$ to calculate the 3MKS. 


\section*{C. System size dependence}
In this appendix, we show the system size dependence of our numerics for some typical cases. 

Figures \ref{Sys_d} (a) and (b) are the system size dependence of $\langle C_{max}\rangle$ for $r=1$ and $p_v$ for $r=1$, where we observe the groundstate $|\psi_g\rangle$ with the parameter sweep (A). We verified that the dependence is entirely much small for $\langle C_{max}\rangle$ and $p_v$ exhibits tiny system size dependence in the extended phase, but no system size dependence in the critical phase. 

Figures \ref{Sys_d} (c) and (d) are the system size dependence of $\langle C_{max}\rangle$ for $r=1$ and $p_v$ for $r=1$, where we observe the groundstate $|\psi_g\rangle$ with the parameter sweep (B). We verified that the dependence is entirely much small for $\langle C_{max}\rangle$ and $p_v$ also exhibits tiny system size dependence in the extended phase, but no system size dependence on the phase boundary. 

Furthermore, figures \ref{Sys_dd} (a) and (d) are the system size dependence of $\langle C^{3MKS}_{max}\rangle$, where we observe the groundstate $|\psi_g\rangle$ with the parameter sweep (A). We verified that the dependence of $\langle C^{3MKS}_{max}\rangle$ is much small around the phase transition point and in the critical phase, while in the extended phase the small system size dependence appears. 
$p^{MKS}_v$ also exhibits a small system size dependence in the critical phase, but no system size dependence around phase transition point.





\begin{thebibliography}{99}
\bibitem{Horodecki_family}
R. Horodecki, P. Horodecki, M. Horodecki and K. Horodecki, Rev. Mod. Phys. {\bf 81}, 865 (2009).

\bibitem{Eisert2010}
J. Eisert, M. Cramer, and M. B. Plenio, Rev. Mod. Phys. {\bf 82}, 277 (2010).

\bibitem{Wen_text}
B. Zeng, X. Chen, D.-L. Zhou, and X.-G. Wen, Quantum Information Meets Quantum Matter: From Quantum Entanglement to Topological Phases of Many-Body Systems (Springer, 2019).

\bibitem{Bell1969}
J.S. Bell, Physics {\bf 1}, 195 (1964).

\bibitem{CHSH1969}
J. F. Clauser, M. A. Horne, A. Shimony, R. A. Holt, Phys. Rev. Lett. {\bf 23}, 880 (1969).

\bibitem{Werner1989}
R. F. Werner, Phys. Rev. A {\bf 40}, 4277 (1989).

\bibitem{Augusiak2015}
R. Augusiak, M. Demianowicz, J. Tura, A. Acín, 
Phys. Rev. Lett. {\bf 115}, 030404 (2015).

\bibitem{Gisin1991}
N. Gisin, Phys. Lett. A {\bf 154}, 201 (1991)

\bibitem{Campbell2010}
S. Campbell and M. Paternostro, Phys. Rev. A {\bf 82}, 042324 (2010).

\bibitem{Justino2012}
L. Justino and T. R. De Oliveira, Phys. Rev. A {\bf 85}, 052128 (2012).

\bibitem{Huang2013}
H. L. Huang, Z. Y. Sun, and B. Wang, Eur. Phys. J. B {\bf 86}, 279 (2013).

\bibitem{Sun2014}
Z. Y. Sun, S. Liu, H. L. Huang, D. Zhang, Y. Y. Wu, 
J. Xu, B. F. Zhan, H. G. Cheng, C. B. Duan, and B. Wang, Phys. Rev. A {\bf 90}, 062129 (2014).

\bibitem{Wen2022}
H. X. Wen, Z. Y. Sun, H. G. Cheng, D. Zhang, and Y. Y. Wu, 
Eur. Phys. J. B {\bf 95}, 95:148 (2022).

\bibitem{Liu2022}
Y. Liu, M. Li, J. Bao, B. Guo, and Z. Sun, Phys. Lett. A {\bf 450}, 128396 (2022).

\bibitem{Lee2020}
D. Lee and W. Son, Entropy {\bf 22}, 1282 (2020).

\bibitem{Getelina2018}
J. C. Getelina, T. R. de Oliveira, and J. A. Hoyos, Phys. Lett. A {\bf 382}, 2799 (2018).

\bibitem{Liang2023}
Z. Liang, J. Bao, L. Shen, B. Guo, and Z. Sun, Phys. Lett. Sect. A {\bf 472}, 128810 (2023).

\bibitem{Oliveira2012}
T. R. de Oliveira, A. Saguia, and M. S. Sarandy, EPL, {\bf 100}, 60004 (2012).

\bibitem{Hatsugai1990}
Y. Hatsugai and M. Kohmoto, Phys. Rev. B {\bf 42}, 8282 (1990).

\bibitem{Han1994}
J. H. Han, D. J. Thouless, H. Hiramoto, and M. Kohmoto, Phys. Rev. B {\bf 50}, 11365 (1994).

\bibitem{Liu2015}
F. Liu, S. Ghosh, and Y. D. Chong, Phys. Rev. B {\bf 91}, 014108 (2015).

\bibitem{Svetlichny1987}
G. Svetlichny, Phys. Rev. D {\bf 35} 3066 (1987).

\bibitem{Mermin1990}
N. D. Mermin, Phys. Rev. Lett. {\bf 65}, 1838 (1990) 

\bibitem{Belinskii}
A. V. Belinskii, D. N. Klyshko, Physics-Uspekhi {\bf 36}, 653 (1993).

\bibitem{Collins2002}
D. Collins, N. Gisin, S. Popescu, D. Roberts, and V. Scarani, Phys. Rev. Lett. {\bf 88}, 1704051 (2002).

\bibitem{Bancal2009}
J. D. Bancal, C. Branciard, N. Gisin, and S. Pironio, Phys. Rev. Lett. {\bf 103}, 090503 (2009).

\bibitem{LSM1961}
E. Lieb, T. Schultz and D. Mattis, Ann. Phys. (New York) {\bf 16}, 407 (1961).

\bibitem{Horodecki1995}
R. Horodecki, P. Horodecki, and M. Horodecki, Phys. Lett. A {\bf 200}, 340 (1995).

\bibitem{Cheong2004}
S. A. Cheong and C. L. Henley, Phys. Rev. B {\bf 69}, 075111 (2004).

\bibitem{come1}
A similar consideration has been used in Refs.~\cite{Justino2012,Getelina2018}. In addition, we also verified the condition by using the numerical exact diagonalization for the model of $H_f$.

\bibitem{Cirelson}
B. S. Cirel'son, Lett. Math. Phys. {\bf 4}, 93 (1980).

\bibitem{Sachdev1999}
S. Sachdev, Quantum Phase Transitions (Cambridge University Press, Cambridge, UK, 1999).

\bibitem{Osborne2002}
T. J. Osborne and M. A. Nielsen, Phys. Rev. A {\bf 66}, 032110 (2002).

\bibitem{Peschel2009}
I. Peschel and V. Eisler, J. Phys. A: Math. Theor. {\bf 42}, 504003 (2009).

\bibitem{Scipy_opt}
\url{https://docs.scipy.org/doc/scipy/reference/generated/scipy.optimize.minimize.html}


\bibitem{Gisin1998}
N. Gisin and H. Bechmann-Pasquinucci, Phys. Lett. A {\bf 246}, 1 (1998).







\end{thebibliography}
\end{document}